# Structural and ferromagnetic properties of an orthorhombic phase of MnBi stabilized with Rh additions

Valentin Taufour,[1, 2, *] Srinivasa Thimmaiah,[2] Stephen March,[1] Scott Saunders,[1] Kewei Sun,[2] Tej Nath Lamichhane,[1, 2] Matthew J. Kramer,[2, 3] Sergey L. Bud'ko,[1, 2] and Paul C. Canfield[1, 2]
[1] *Department of Physics and Astronomy, Iowa State University, Ames, Iowa 50011, USA*
[2] *The Ames Laboratory, U.S. Department of Energy, Iowa State University, Ames, Iowa 50011, USA*
[3] *Department of Materials Science and Engineering, Iowa State University, Ames, Iowa 50011, USA*

The article addresses the possibility of alloy elements in MnBi which may modify the thermodynamic stability of the NiAs-type structure without significantly degrading the magnetic properties. The addition of small amounts of Rh and Mn provides an improvement in the thermal stability with some degradation of the magnetic properties. The small amounts of Rh and Mn additions in MnBi stabilize an orthorhombic phase whose structural and magnetic properties are closely related to the ones of the previously reported high-temperature phase of MnBi (HT MnBi). To date, the properties of the HT MnBi, which is stable between 613 and 719 K, have not been studied in detail because of its transformation to the stable low-temperature MnBi (LT MnBi), making measurements near and below its Curie temperature difficult. The Rh-stabilized MnBi with chemical formula $Mn_{1.0625-x}Rh_xBi$ [$x = 0.02(1)$] adopts a new superstructure of the NiAs/$Ni_2In$ structure family. It is ferromagnetic below a Curie temperature of 416 K. The critical exponents of the ferromagnetic transition are not of the mean-field type but are closer to those associated with the Ising model in three dimensions. The magnetic anisotropy is uniaxial; the anisotropy energy is rather large, and it does not increase when raising the temperature, contrary to what happens in LT MnBi. The saturation magnetization is approximately 3 $\mu_B$/f.u. at low temperatures. While this exact composition may not be application ready, it does show that alloying is a viable route to modifying the stability of this class of rare-earth-free magnet alloys.

## I. INTRODUCTION

The compound MnBi has attracted attention because of its remarkable magnetic and structural properties. MnBi is a rare Mn-based ferromagnet with a large magnetic anisotropy, which uniquely increases with increasing temperature, and is potentially interesting for rare-earth-free permanent magnet applications. As such, MnBi is considered as a promising hard phase for exchange coupling nanocomposite magnets which could find unique green-energy applications, such as high-performance electric motors, low-energy-consumption power electronics, and wind power generators [1–3]. The MnBi family of compounds has been actively pursued by a number of groups in China, Europe and the United States as a replacement for rare-earth-based magnets. One of the more significant challenges is the moderately low decomposition temperature of 628 K. In recent years, several processing and chemical substitution studies have been undertaken to improve the properties of MnBi and gain further insight into the origin of its intriguing characteristics [2, 4–9].

The binary MnBi system contains two different compounds. The low-temperature (LT) phase which has the hexagonal NiAs crystal structure [space group $P6_3/mmc$ (194) [10]], is stable below 628 K [11], and the high-temperature (HT) phase, which has a distorted NiAs crystal structure (no consensus on the space group [11–14]), is stable between 613 and 719 K, above which it decomposes peritectically into Mn plus liquid [11]. The composition of the HT phase is $Mn_{1.08}Bi$.

LT MnBi is ferromagnetic up to the coupled structural and magnetic phase transitions at 628 K, above which the HT $Mn_{1.08}Bi$ is in a paramagnetic state. In rapidly quenched HT $Mn_{1.08}Bi$, a ferromagnetic state is found with a Curie temperature of about 440 K [12, 15]. At room temperature, quenched HT $Mn_{1.08}Bi$ is thermally unstable and transforms slowly back into the LT MnBi with a time constant of about two years [16]. Near the Curie temperature, the time constant is reduced to below a few minutes, so that the magnetic properties of the quenched HT $Mn_{1.08}Bi$ cannot be measured directly near or above $T_C$ [15]. The transformation proceeds via a metastable third phase, the structure of which is not yet definitely identified [17, 18].

As result of the peritectic decomposition at high temperatures, when LT MnBi is prepared by using conventional induction or arc-melting casting methods, additional Mn as well as postcasting annealing is required to form the desired LT phase. The HT $Mn_{1.08}Bi$ is often obtained during the preparation of LT MnBi by various metallurgical methods [9]. It is therefore essential to elucidate its structural and magnetic properties as well as to investigate the circumstances under which this phase can become more or less stable.

* taufour@ameslab.gov

Whereas band-structure calculations suggest that doping the LT MnBi with Rh increases the magnetocrystalline anisotropy and the coercivity [7], in this study, we observe that the addition of Rh to MnBi is effective in stabilizing an orthorhombic phase relative to the LT phase, as previously reported for thin films and bulk powders [19] and similar to the effect of Ru additions [19]. Analogous effects are observed with Ti on MnBi thin films [18, 19]. We report on the crystal structure and magnetic properties of Rh-stabilized single crystals of orthorhombic $Mn_{1.0625-x}Rh_xBi$ with $x = 0.017(1)$, which we write $Mn_{1.05}Rh_{0.02}Bi$. In the absence of a consensus on the crystal structure of HT $Mn_{1.08}Bi$ [11–14], it is unclear if the Rh-stabilized orthorhombic phase and the HT phase have precisely the same crystal structure, but we find that they have similar magnetic properties. The magnetic properties of Rh-stabilized orthorhombic MnBi have been studied without noticing instability of the crystal structure over the time of this study. The Curie temperature is 416 K with critical exponents close to the Ising model in three dimensions. The saturation magnetization is approximately 3 $\mu_B$/f.u at 50 K, and the magnetocrystalline anisotropy energy of approximately 4 MJ/m$^3$ is much larger than for the LT MnBi. The magnetocrystalline anisotropy energy does not increase when raising the temperature, contrary to what happens in LT MnBi. The reduction of the Curie temperature and of the magnetization compared with LT MnBi shows that this exact composition may not be application ready, but it does show that alloying is a viable route to modifying the stability of this class of rare-earth-free magnet alloys.

## II. EXPERIMENTAL TECHNIQUES

### A. Single crystal growth

Single crystals of Rh-doped MnBi were grown by standard solution growth methods [20] using an approximately 5 g mixture of Rh, Mn, and Bi placed into a 2 mL alumina crucible with a molar ratio of Mn:Rh:Bi $= (10 - x) : x : 90$ $(0 \leqslant x \leqslant 2.0)$. The crucible with the starting elements was sealed in a fused-silica ampoule under a partial argon atmosphere, which was then heated to 1150 °C, held at 1150 °C for 5 h, cooled to 390 °C over 5 h, held at 390 °C for 2 h, and then cooled at a rate of approximately 1 °C per hour to 280 °C. Once at a stable 280 °C, the solution was decanted with the assistance of a centrifuge [20, 21].

Under these temperature conditions, MnBi crystals are expected to grow in the LT phase [11]. We found that this is true for $x = 0$, and we obtained MnBi single crystals with similar shape and size as other recent reports [22]. But growths with small amounts of Rh additions revealed that the Mn-Rh-Bi system produces well-formed hexagonal plates, uncharacteristic of the typical small hexagonal box morphology of the pure MnBi binary system (see Fig. 1). Additionally, the crystals produced at the two different, initial concentrations ($x = 1.0$ and 2.0) were similar in size, indicating morphology is influenced by the presence of Rh but not a variable of Rh for $x = 1.0$ and 2.0. The Rh concentration was further adjusted by using $Mn_{10-x}Rh_xBi_{90}$ $(x = 0.10, 0.20, 0.25, 0.30, 0.40, 0.5, 0.75)$ following the same growth parameters as before. Even as $x$ was reduced to 0.10, large platelike crystals were produced: the addition of Rh into the Mn-Bi system produces a dramatic change in morphology, even for small Rh concentrations. This morphological change was our first indication that there might well be a crystallographic, and possibly even magnetic, change in the resulting Rh-doped samples.

(a) (b)

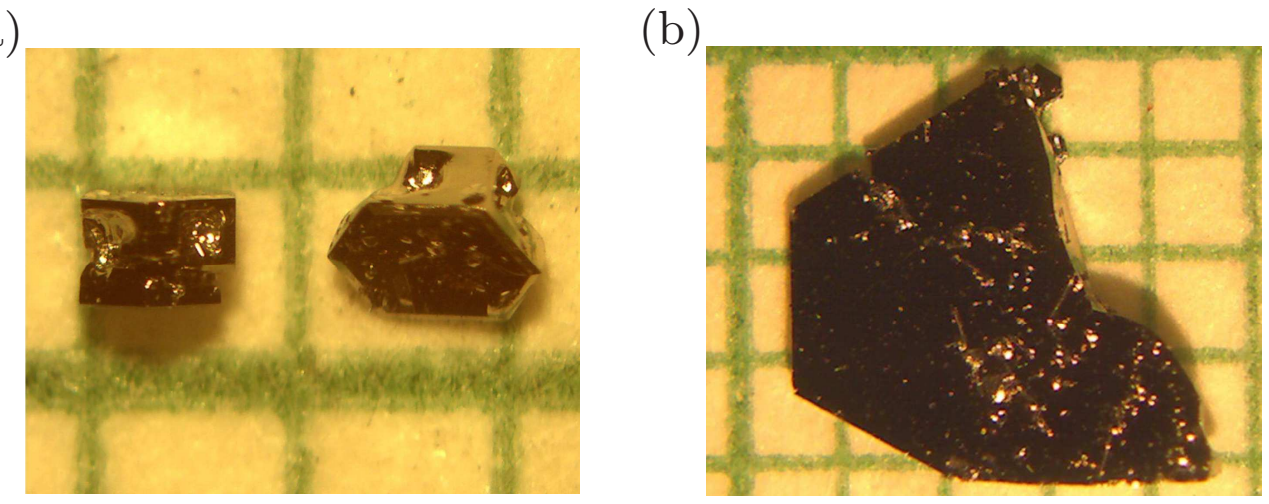

FIG. 1. (Color online) Single crystal samples grown from melts with initial compositions of $Mn_{10-x}Rh_xBi_{90}$ for (a) $x = 0$ (MnBi), (b) $x = 0.5$ on a 1 mm grid. The Rh addition causes a dramatic change in morphology and size. This was the first hint that a new crystal structure may have occurred.

### B. X-ray diffraction, elemental analysis, and transmission electron microscopy

For single crystal diffraction studies, crystals were isolated from the growth with $x = 0.3$. Several crystals with typical sizes of $0.08 \times 0.05 \times 0.03$ mm$^3$ were measured from the same batch to verify the consistency of the Rh content in the structure. Diffraction intensities were collected at room temperature on a SMART APEX II diffractometer equipped with a CCD area detector using graphite monochromated MoK radiation. The crystal-to-detector distance was 6.0 cm, and the irradiation time was 25–30 s/frame. Data-collection strategies were obtained from an algorithm in the program COSMO as part of the APEX II software package [23]. Additional details are given in the Appendix.

Elemental analysis of the samples was performed using wavelength-dispersive x-ray spectroscopy (WDS) in a JEOL JXA-8200 electron probe microanalyzer. The WDS data were collected from multiple points on the samples. Impurity traces of $Mn_3Rh$, $MnBiO_2$ and slightly oxidized Bi were also detected on the small droplets of residual melt on the surface of the samples [e.g., see lower right side of crystal in Fig. 1(b)].

The transmission electron microscopy samples (1 mm long and 1 $\mu$m wide) were prepared using mechanical wedge polishing followed by Ar ion milling in liquid nitrogen temperature. The specimen was thinned in the range from 20 to 50 nm to insure electron transparency. The transmission electron micrographs were taken by Tecnai F20 transmission electron microscope operated in 200 kV.

### C. Magnetic measurements

Magnetization measurements were done by using a Vibrating Sample Magnetometer (VSM) in a cryogen-free Physical Property Measurement System (Versalab, Quantum Design) with a magnetic field up to 3 T in the temperature range of 50–350 K using the standard option and 300–500 K using the oven option. An alumina cement (Zircar) was used to hold the sample on the heater stick for the high-temperature measurements. The demagnetization factors were calculated using Ref. [24], and the internal field $H_{\rm int}$ was calculated from the relation $H_{\rm int} = H - NM$. Several crystals from two different batches were measured to verify the consistency of the results.

## III. RESULTS AND DISCUSSION

### A. Crystal structure, elemental analysis, and transmission electron microscopy

Single crystal diffraction experiments reveal that the Rh-doped $Mn_{1+x}Bi$ adopts several new structure types in the orthorhombic crystal system depending on the Rh content. Here we report on the samples with a refined composition of $Mn_{1.0625-x}Rh_xBi$ with $x \approx 0.017$, which we denote $Mn_{1.05}Rh_{0.02}Bi$ (see the Appendix). Another slightly different superstructure is obtained at a higher Rh composition of $Mn_{1.04(1)}Rh_{0.03(1)}Bi$, and a detailed description of these closely related structures will be presented elsewhere [25].

According to the refinement on a single crystal, Rh atoms partially replace one set of Mn atoms ($8a$ site) in the crystal structure. The refined composition of $Mn_{1.05(1)}Rh_{0.02(1)}Bi$ agrees well with the results of WDS analyses where the stoichiometry is found to be $Mn_{1.02(1)}Rh_{0.03(1)}Bi$. For the rest of this work, we use the stoichiometry determined from the single crystal X-ray refinement. This phase crystallizes in an orthorhombic structure (space group $Fdd2$) with lattice parameters $a = 8.683(3)$ Å, $b = 47.704(8)$ Å, and $c = 15.021(3)$ Åand the unit cell volume of 6222(3) Å$^3$ ($Z = 128$). The structure of $Mn_{1.05(1)}Rh_{0.02(1)}Bi$ is a new superstructure of the NiAs/$Ni_2In$ family of structure and is closely related to LT MnBi, which adopts the hexagonal NiAs-type structure [$P6_3/mmc$, $a_h = 4.2909(1)$ Å, $c_h = 6.1233(9)$ Å] [8, 10]. The lattice parameters of the $Mn_{1.05}Rh_{0.02}Bi$ structure can be expressed with respect to LT MnBi as follows: $a_o \approx 2a_h$, $b_o \approx 8c_h$, and $c_o \approx 2\sqrt{3}a_h$. With this structure, the $b$-axis direction corresponds to the $c$ axis of LT MnBi, and it is the direction perpendicular to the plate surface of the single crystals [Fig. 1(b)].

It is unclear if the orthorhombic $Mn_{1.05}Rh_{0.02}Bi$ and the HT $Mn_{1.08}Bi$ have the same crystal structure, because there is no consensus on the crystal structure of HT $Mn_{1.08}Bi$ [11–14]. The crystal structure of rapidly quenched HT $Mn_{1.08}Bi$ was proposed to be a large orthorhombic supercell of dimensions $a = 11.94$ Å, $b = 8.7$ Å, and $c = 7.52$ Å which was derived from the NiAs-type cell by making $a = 2c_h$, $b = 2a_h$ and $c = \sqrt{3}a_h$ [11, 12]. In Ref. [12], the symmetry of the cell was believed to be more likely monoclinic than orthorhombic, but the direction and the magnitude of the atomic displacements could not be obtained from powder data. In Ref. [13], the orthorhombic crystal structure with the space group $P222_1$ with $a = 4.334(2)$ Å, $b = 7.505(4)$ Å, and $c = 5.959(7)$ Å was proposed, but later, from the same data, the space group $Pmma$ was proposed [14].

Figure 2(a) shows the crystalline morphology of the crystal. The corresponding selected area electron diffraction pattern was obtained along the $b$ axis of the crystal in Fig. 2(b). The transmission electron micrographs demonstrate a pretty uniform structure with a domain size of hundreds of nanometers. Moiré fringes are well observed, indicating a misorientation at the domain boundaries. High-resolution transmission electron microscopy taken at the domain boundaries in Fig. 2(c) highlights that the structure is composed of three interpenetrating [010] zones with 120° rotation to each other. This results in a pseudohexagonal symmetry which is observed in the morphology of the large as-grown crystals [see Fig. 1(b)].

### B. Magnetic properties

$Mn_{1.05}Rh_{0.02}Bi$ is found to be ferromagnetic. First, we discuss its uniaxial magnetic anisotropy. Then, we describe how we determine the Curie temperature $T_C \approx 416$ K and the critical behavior.

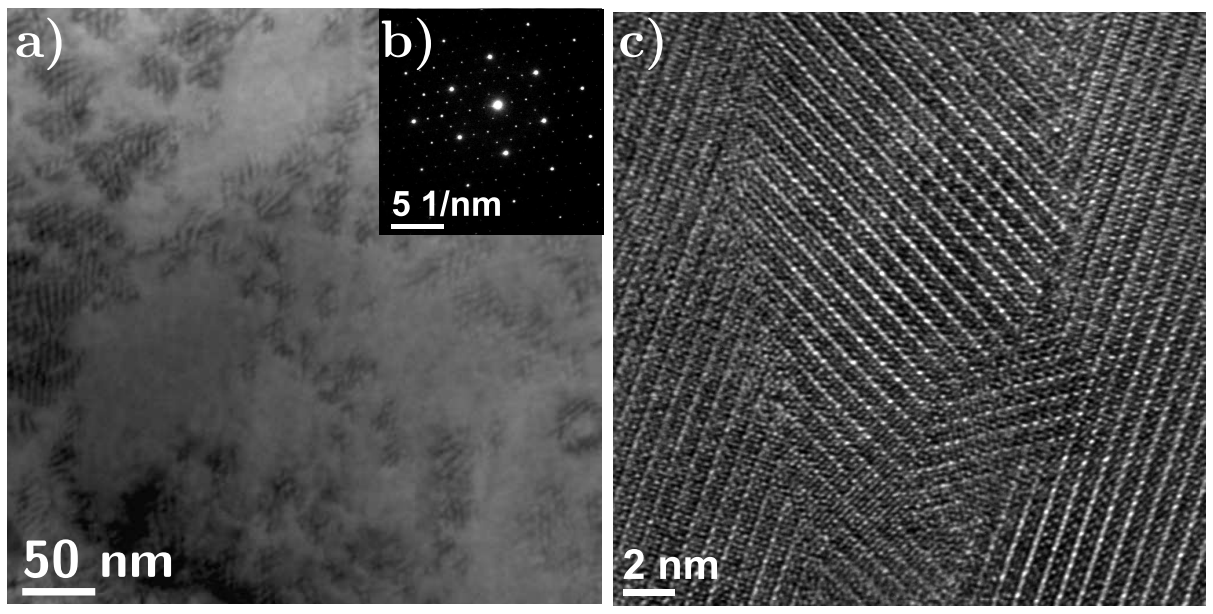

FIG. 2. (a) Transmission electron micrographs showing the morphology of the crystal and (b) corresponding electron diffraction demonstrating the pseudohexagonal symmetry. (c) High-resolution transmission electron microscopy showing the lattice fringes with 120° rotation.

The temperature dependence of the magnetization is shown in Fig. 3. The magnetization along the $b$ axis [per-pendicular to the plate-like surface shown in Fig. 1(b)] rapidly rises upon cooling around 430 K.

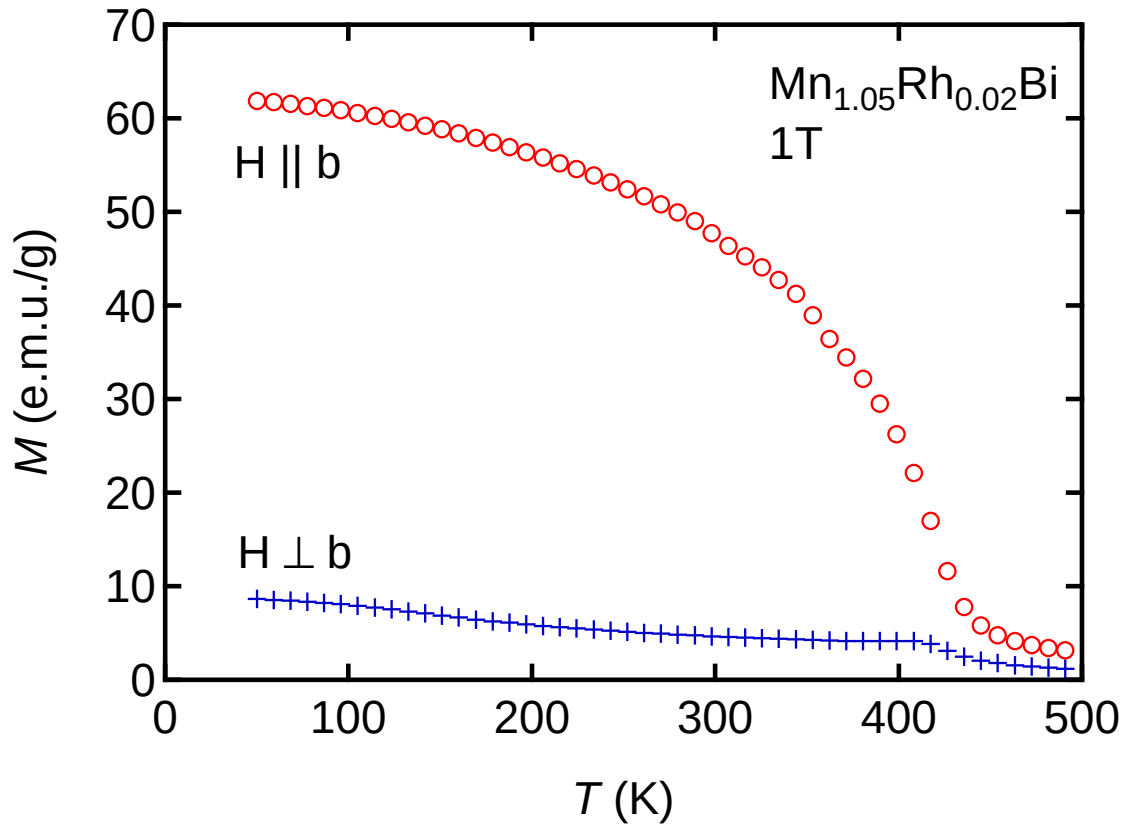

FIG. 3. (Color online) Temperature dependence of the magnetization of $Mn_{1.05}Rh_{0.02}Bi$ along the $b$ axis and perpendicular to the $b$ axis. The $b$ axis is perpendicular to the platelike surface shown in Fig. 1(b).

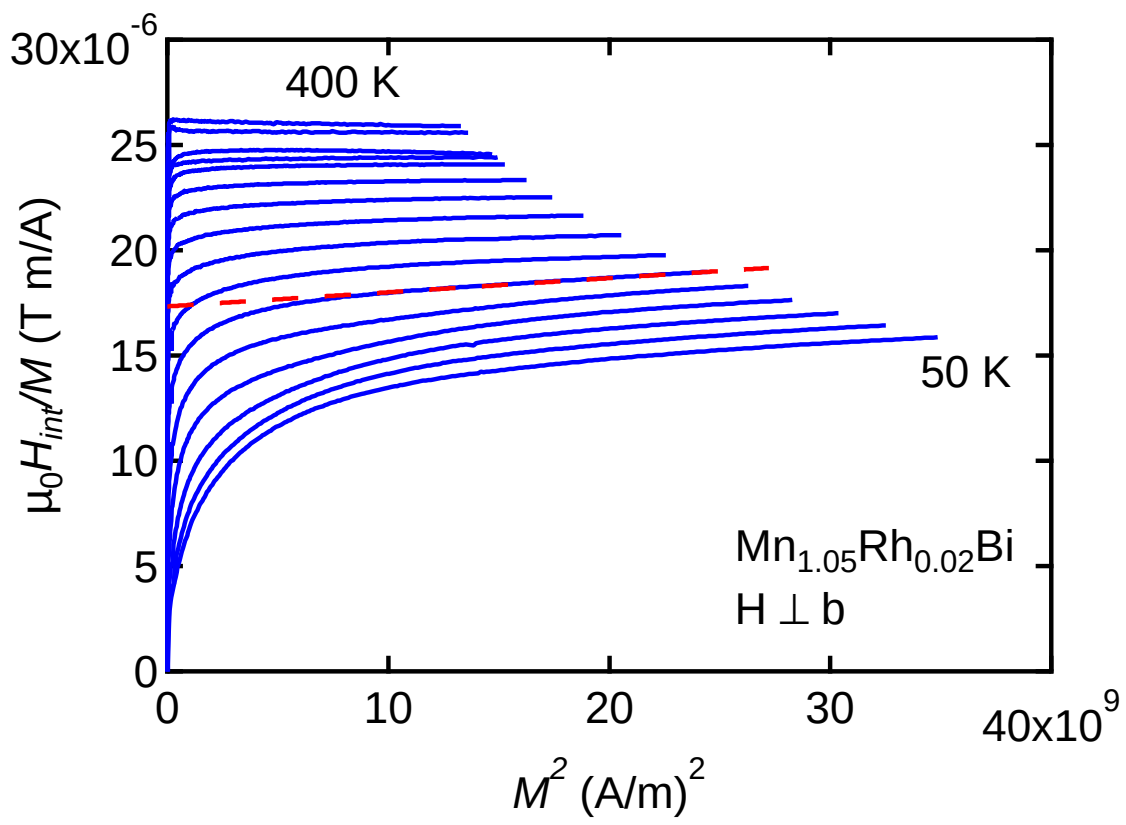

FIG. 5. (Color online) $H/M$ versus $M^2$ for the magnetic field applied perpendicular to the $b$ axis at different temperatures from 50 to 400 K in 25 K steps. The dashed line is an example of a linear fit used to determine the anisotropy constants $K_1$ and $K_2$ from the Sucksmith method [26, 27].

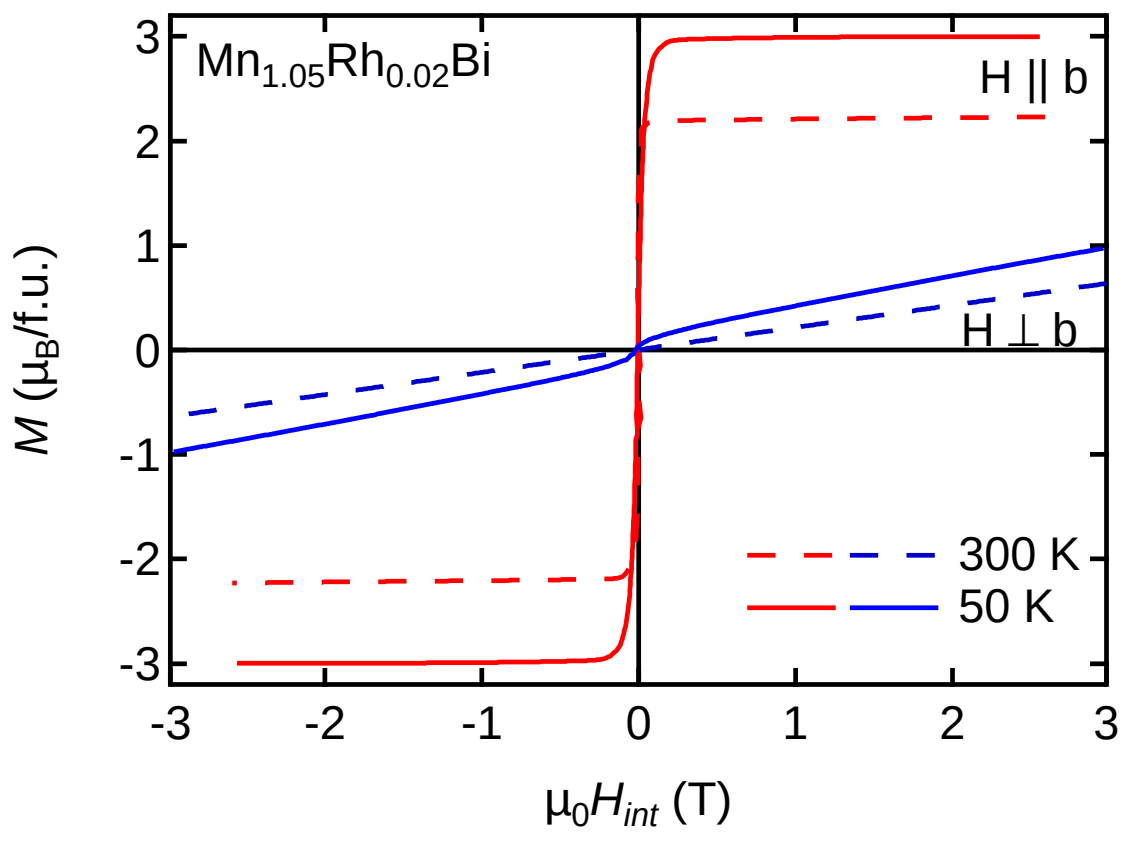

FIG. 4. (Color online) Magnetic-field dependence of the magnetization at 50 and 300 K for the field applied along the $b$ axis and perpendicular to the $b$ axis.

Figure 4 shows the magnetic-field dependence of the magnetization at 50 and 300 K for the field applied parallel and perpendicular to the $b$ axis. The saturation magnetization at 50 K is 3 $\mu_B$/f.u. (2.86 $\mu_B$/Mn) which is very close to the value of 3.07 $\mu_B$/Mn observed in rapidly quenched HT $Mn_{1.08}Bi$ at 4 K [15]. A large magnetic anisotropy is clearly visible, with the $b$ axis being the direction of easy magnetization. No magnetic hysteresis can be observed, which is not unusual for single crystals, because of very small domain pinning.

Although $Mn_{1.05}Rh_{0.02}Bi$ has an orthorhombic structure, the $a-c$ plane has a pseudohexagonal arrangement (see Fig. 10 in the Appendix). In addition, as revealed by our transmission electron microscopy, large crystals are composed of interpenetrating [010] zones with 120° rotation to each other, which results in a pseudohexagonal symmetry. Therefore, the crystals can be considered as uniaxial magnets. In order to determine the anisotropy constant, we use the Sucksmith method [26, 27], which can be used for uniaxial systems. In this method, the first- and second-order uniaxial anisotropy constants $K_1$ and $K_2$ can be determined by plotting $H/M_\perp$ versus ${M_\perp}^2$ (see Fig. 5) and by fitting the linear part by

$$\frac{H}{M_\perp} = \frac{2K_1}{{M_s}^2} + \frac{4K_2}{{M_s}^4}{M_\perp}^2 \quad (1)$$

$M_s$ can be obtained from the magnetization curves along the easy axis, as shown in Fig. 4. The obtained values of $K_1$ and $K_2$ are shown as a function of temperature in Fig. 6. The total anisotropy energy $K_1 + K_2$ is very similar to the one of the quenched HT $Mn_{1.08}Bi$ reported in Ref. [15]. The anisotropy energy is much larger than for the LT MnBi. The temperature dependence shows a decrease with increasing temperature which is more conventional than for the LT MnBi.

In order to obtain a precise Curie temperature, we first try to use Arrott plots [28] by plotting isotherms of $M^2$ versus $H/M$ as shown in Fig. 7(a). Since all isotherms have a positive slope, the paramagnetic-to-ferromagnetic transition is of the second order [29]. If one would consider only the low-field region and force a line to go through the origin, a value of $T_C$ around 418.5 K would be obtained. However, no isotherm goes linearly through the origin, which implies that the transition cannot be described by a mean-field model with critical exponents $\beta = 0.5$, $\gamma = 1$, and $\delta = 3$. The critical exponents are defined so that the magnetization $M$ has a singular part proportional to $|(T - T_c)/T_c|^\beta$ and the susceptibility $\chi$ has a singular part proportional to $|(T - T_c)/T_c|^{-\gamma}$. In addition, the equation of state at $T_c$ is $M \propto H^{1/\delta}$ [30]. A modified Arrott plot of $M^{1/\beta}$ versus $(H/M)^\gamma$ is shown in Fig. 7(b) for $\beta = 0.32$ and $\gamma = 1.43$. We can see that this choice of critical exponents shows an isotherm going linearly through the origin. Here we describe how these values for the critical exponents are obtained.

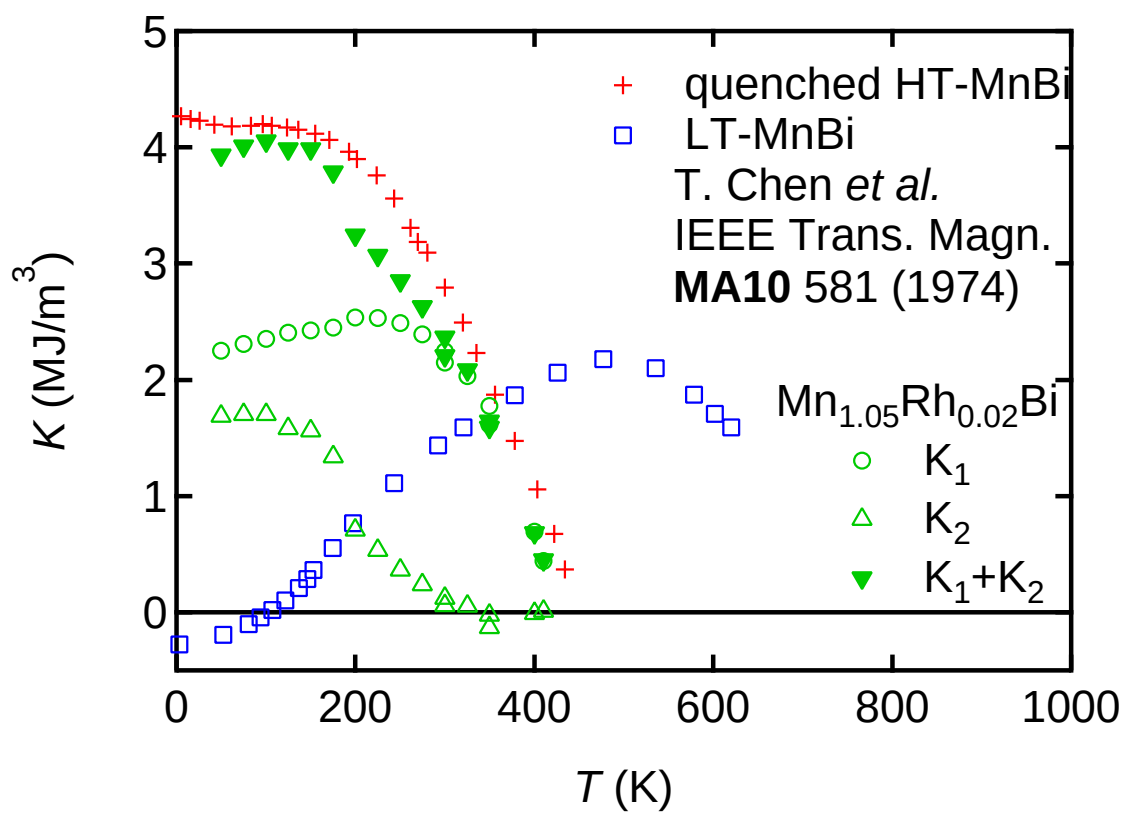

FIG. 6. (Color online) Temperature dependence of the anisotropy energy constants $K_1$ and $K_2$ and total anisotropy energy $K_1 + K_2$ of $Mn_{1.05}Rh_{0.02}Bi$. The anisotropy energy of the quenched HT MnBi and of the LT MnBi is shown for comparison [15].

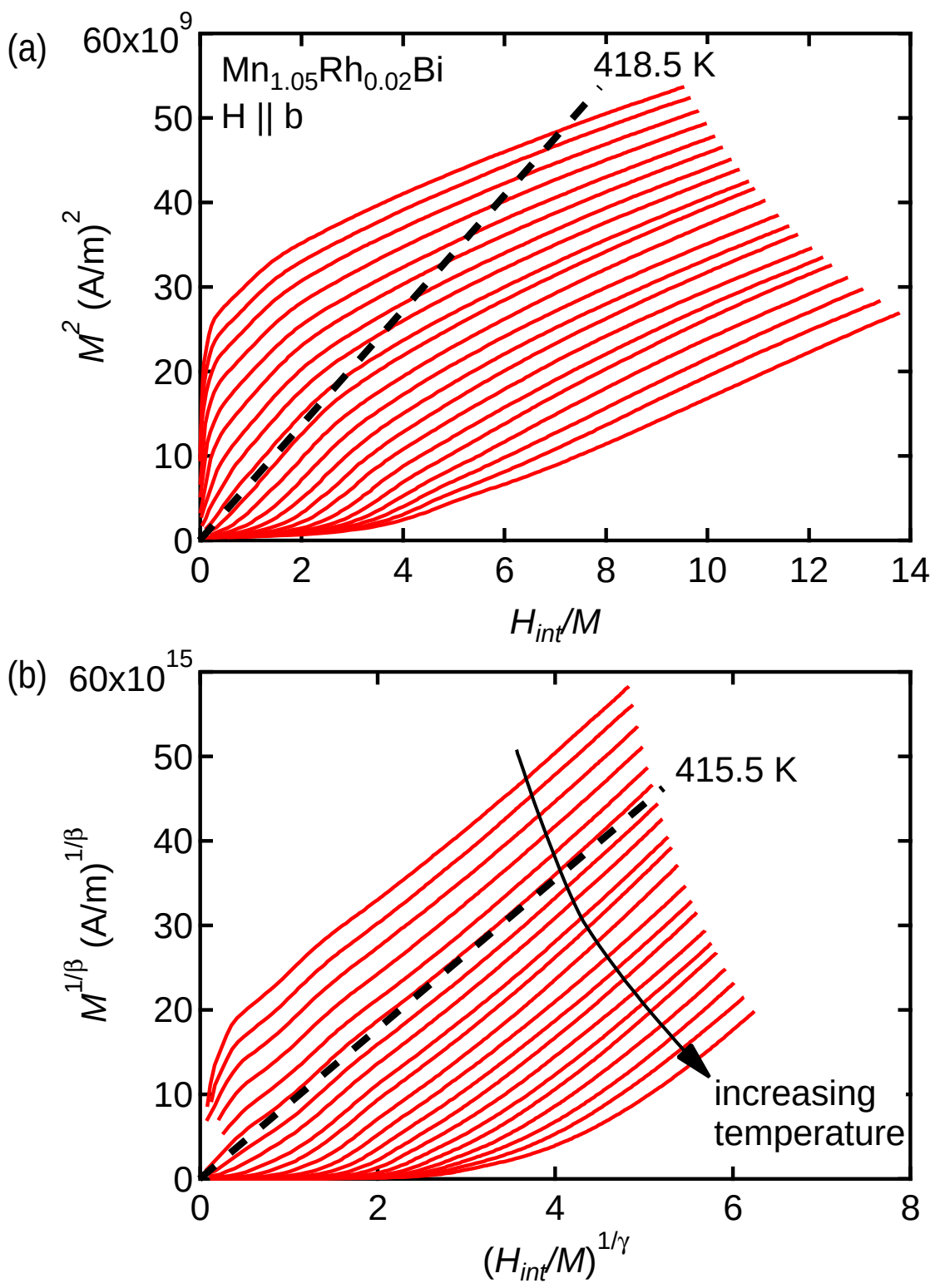

FIG. 7. (Color online) (a) Arrott plots in the form of $M^2$ versus $H/M$ measured at different temperatures from 410 to 430 K in 1 K steps. The field is applied along the $b$ axis. (b) Modified Arrott plot in the form of $M^{1/\beta}$ versus $(H/M)^{1/\gamma}$ with $\beta = 0.32$ and $\gamma = 1.43$.

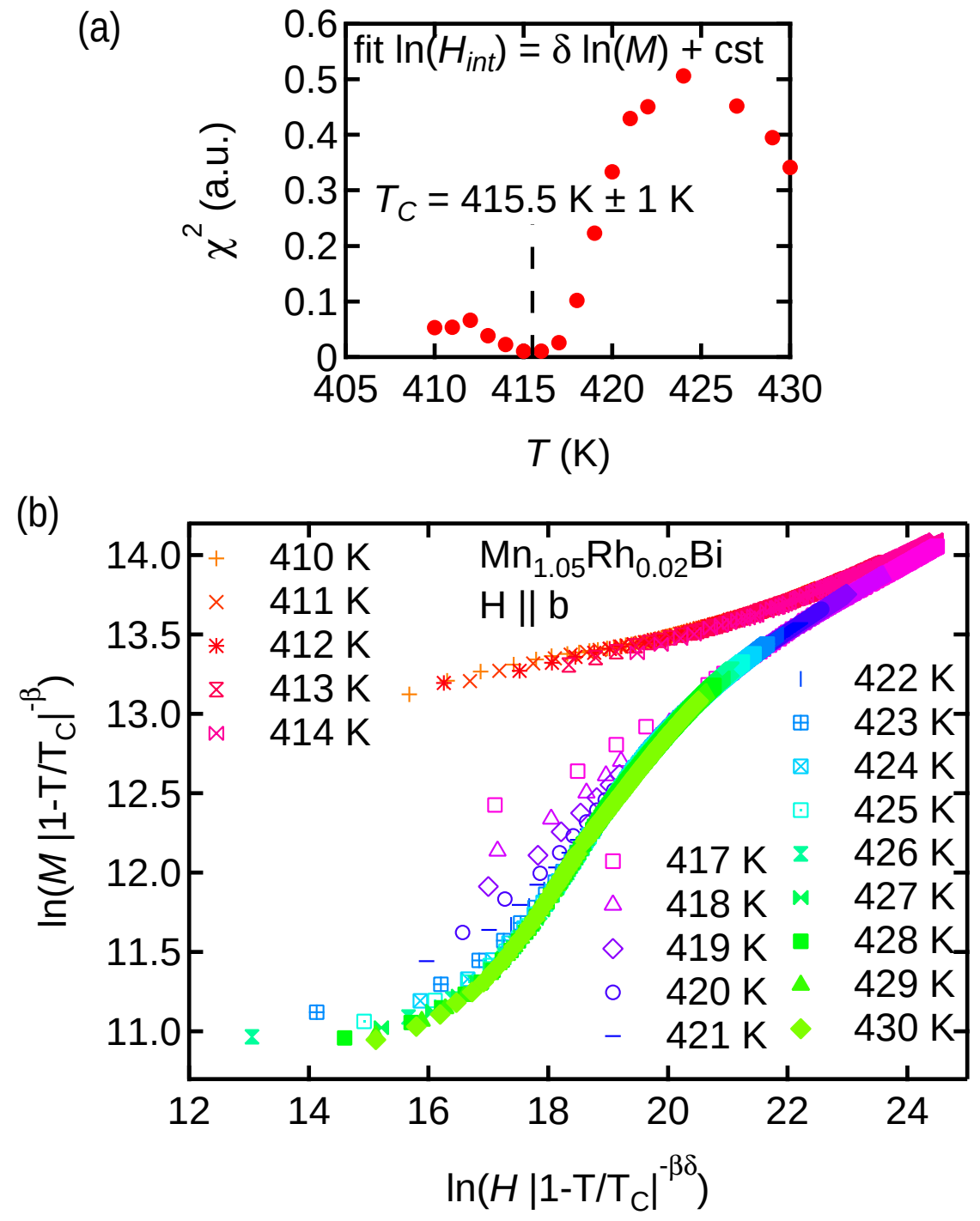

FIG. 8. (Color online) (a) Temperature dependence of the goodness of the fit $\ln(H) = \delta\ln(M) + k$ ($k$ is a constant) in the form of chi squared. The best fit is obtained around $T_C = 415.5$ K for which $\delta = 5.48$. (b) Log of reduced magnetization $\ln\left(M\,|1 - T/T_C|^{-\beta}\right)$ versus log of reduced field $\ln\left(H\,|1 - T/T_C|^{-\beta\delta}\right)$. $T_C = 415.5$ K and $\delta = 5.48$ are fixed from (a). The only variable is the critical exponent $\beta$, which is adjusted until all the isotherms merge into only two curves, corresponding to below and above $T_C$. We obtain $\beta = 0.32$.

The equation of state at $T_C$ can be written in the form $\ln(H) = \delta\ln(M) + k$, where $k$ is a constant. We performed linear fits of $\ln(H)$ versus $\ln(M)$ for all isotherms. The goodness of the fit is shown in Fig. 8(a) in the form of chi squared. We obtain the best fit at $T_C = 415.5$ K. At this temperature, we also obtain $\delta = 5.48$. With these fixed values of $T_C$ and $\delta$, we then plot the scaling hypothesis [31–36] in the form $\ln\left(M\,|1 - T/T_C|^{-\beta}\right)$ versus $\ln\left(H\,|1 - T/T_C|^{-\beta\delta}\right)$ [see Fig. 8(b)]. In the scaling hypothesis, we adjust the only parameter $\beta$ until all the isotherms merge into only two curves corresponding to data below and above $T_C$. With this procedure, we obtain $\beta = 0.32$ from which we calculate $\gamma = 1.43$ by using the Widom relation [37] ($\delta = 1 + \gamma/\beta$). Then, we can make the modified Arrott plot shown in Fig. 7(b).

The obtained critical exponents ($\delta = 5.48$, $\beta = 0.32$, and $\gamma = 1.43$) are closer to the Ising model in dimension $d = 3$ [38] ($\delta = 5$, $\beta = 5/16 \approx 0.31$, and $\gamma = 1.25$) than to the three-dimensional Heisenberg universality class ($\delta \approx 4.78$, $\beta \approx 0.37$, and $\gamma \approx 1.40$) [39]. An Ising class is consistent with the rather large uniaxial anisotropy that we discussed previously.

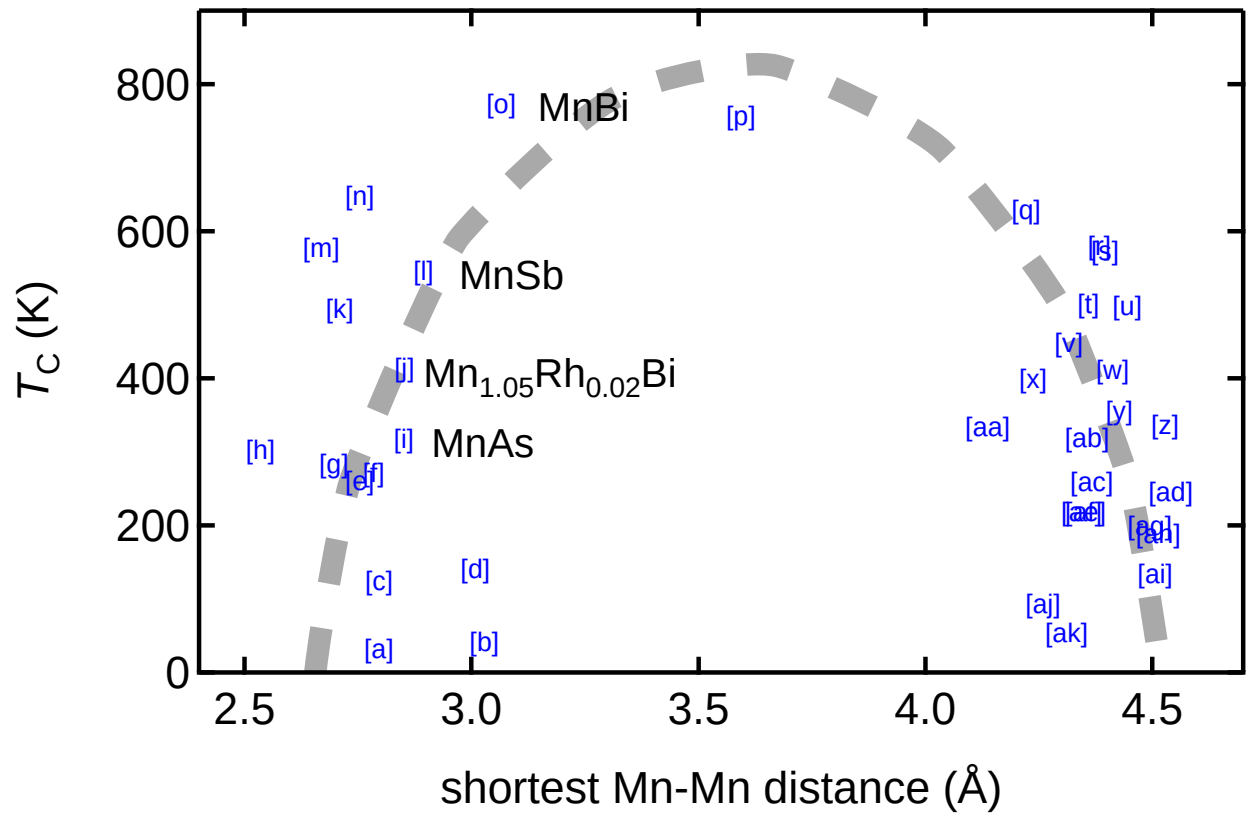

FIG. 9. (Color online) Schematic representation of the Castelliz-Kanomata empirical interaction curve [40, 41] showing the Curie temperature as a function of the interatomic Mn-Mn distance in various Mn-based compounds: [a] MnSi, [b] $Mn_3O_4$, [c] $Mn_4As_3$, [d] $MnB_2$, [e] $Mn_{1.75}Sn$, [f] $Mn_3As_2$, [g] MnP, [h] $Mn_5Ge_3$, [i] MnAs, [j] $Mn_{1.05}Rh_{0.02}Bi$, [k] $MnPt_3$, [l] $Mn_{1.1}Sb$, [m] MnB, [n] MnAl, [o] MnBi, [p] $MnNi_3$, [q] $AlCu_2Mn$, [r] MnPtSb, [s] $Cu_2InMn$, [t] $Cu_2MnSn$, [u] MnPdSb, [v] $GeMnRh_2$, [w] $MnRh_2Sn$, [x] AlIrMn, [y] MnPtSn, [z] $MnPbRh_2$, [aa] $MnRh_2Sb$, [ab] MnRhSb, [ac] $GeMnPd_2$, [ad] $MnPd_2Sb$, [ae] GaMnPt, [af] GaMnRu, [ag] $AlAu_2Mn$, [ah] $MnPd_2Sn$, [ai] AuMnSb, [aj] $AlMnRh_2$, [ak] CuMnSb.

In general, the ferromagnetism of Mn-based compounds can be described within the framework of density-functional theory [7, 42–45]. For the present crystal structure, with 264 atoms per unit cell, the computational cost can be significant. At the gross level, it is possible to relate $Mn_{1.05}Rh_{0.02}Bi$ with the other Mn-based materials following the empirical interaction curve from Castelliz [40]. Such a curve is shown in Fig. 9, where the Curie temperatures of several Mn-based compounds are plotted as a function of the shortest Mn-Mn distance. Materials containing other magnetic elements such as rare earths are not included. Compounds located on the right side of the curve have a positive pressure derivative of the Curie temperature ($dT_C/dp > 0$), whereas $dT_C/dp < 0$ for compounds on the left side [42]. We can see that $Mn_{1.05}Rh_{0.02}Bi$ is located on this curve between compounds with the NiAs-type structure such as LT MnBi, MnSb and MnAs.

In summary, we have studied the structural and magnetic properties of a Rh-stabilized phase of MnBi: orthorhombic $Mn_{1.05}Rh_{0.02}Bi$. We find that this material is ferromagnetic below $T_C = 416$ K. The critical exponents of the ferromagnetic transition are not of the mean-field type but are closer to the Ising model in three dimensions. The uniaxial anisotropy is rather large and is very close to the one reported for the quenched HT MnBi phase. Our work shows that the addition of Rh in MnBi induces a structural change which must be considered when using doping to increase the magnetocrystalline anisotropy and the coercivity of this material. In addition to the structural properties, the magnetic properties of the orthorhombic $Mn_{1.05}Rh_{0.02}Bi$ are distinguishable from the LT MnBi and resemble more the ones of the HT MnBi.

We thank G. J. Miller, A. Jesche, D. Finnemore, T. Kong, A. Böhmer, and W. E. Straszheim for useful discussions. The research was supported by the Critical Materials Institute, an Energy Innovation Hub funded by the U.S. Department of Energy, Office of Energy Efficiency and Renewable Energy, Advanced Manufacturing Office. This work was also supported by the Office of Basic Energy Sciences, Materials Sciences Division, U.S. DOE. The microscopy was performed under funding from the U.S. DOE's Advanced Research Projects Agency-Energy under contract No. 11/CJ000/09/03. This work was performed at the Ames Laboratory, operated for DOE by Iowa State University under Contract No. DE-AC02-07CH11358.

## Appendix: CRYSTAL STRUCTURE

Structure solutions were obtained via the SHELXT program (Intrinsic Phasing) as implemented in the SHELXTL [46] package, and subsequent structural refinements were performed by a full-matrix least-squares procedure on $|F|^2$ for all data using SHELXL [46]. The final stages of refinements were carried out by using anisotropic displacement parameters on all atoms. Details concerning the structure refinement and atomic coordinates of $Mn_{1.05(1)}Rh_{0.02(1)}Bi$ are tabulated in Tables I and II, respectively.

In LT MnBi, Bi atoms form a hexagonally close packed array of atoms ($c/a$ = 1.427, less than the ideal axial ratio of 1.633) and Mn atoms occupy all octahedral holes. In NiAs-type structures, the trigonal bipyramidal (tbp) sites are completely empty, but these sites are completely filled in the $Ni_2In$-structure type. The formula of the new phase can be expressed in terms of ordering of atoms in the structure as $[Mn^{oct}Bi]_{0.5}[\{Mn_{3/4}\square_{1/4}\}^{oct}(MnRh)^{tbp}_{3/8}Bi]_{0.5}$ ($\square$ = vacancies). The ordered partial occupancy of the tbp site results in various commensurate or incommensurately modulated [47] superstructures of the NiAs/$Ni_2In$ forms [48, 49], which are frequently observed in Mn-Sn [50], Co-In [51], Co-Sn [52, 53], and Ni-Sn [54] systems.

An initial examination of the decoration of the atoms in the unit cell of orthorhombic $Mn_{1.05(1)}Rh_{0.02(1)}Bi$ shows its close relationship to hexagonal LT MnBi, as depicted in Fig. 10. The Bi atoms arrange in more or less a hexagonal close-packed manner and the Mn atoms occupy most of the octahedral voids. However, some of the trigonal bipyramidal (tbp) holes are occupied by a mixture of Mn and Rh atoms in certain Bi hexagonal close-packed planes. To understand the slight enrichment of Mn and Rh atoms, which deviates from the stoichiometric composition MnBi, the structure is decomposed into several atomic layers stacked along the $b$ direction. According to Fig. 10, the observed layer sequence of "Bi" (layers $A$, $A'$, or $B$) and "Mn" (layers $c$ or $c'$) atoms in the $Mn_{1.05(1)}Rh_{0.02(1)}Bi$ structure is $\cdots AcBc'A'c'BcA\cdots$, and the composition of each layer is given for comparison. The Bi content per unit cell within layers $A$, $B$, and $A'$ are the same; however, differences arise due to filling of the tbp holes by Mn and Rh atoms. The tbp holes are partially filled by Mn/Rh and Mn atoms in layers $A$ and $B$, respectively, but remain completely empty in the $A'$ layer. Coinciding with the partial occupation of tbp sites in layers $A$ and $B$, layer $c$ reveals the occurrence of Mn vacancies; i.e., 25% of the octahedral holes are empty. Thus, the formation of the superstructure from LT MnBi (NiAs-type) is the result of ordering vacancies in the octahedral holes and partial filling of tbp holes in the NiAs-type aristotype.

TABLE I. Crystallographic data and refinement parameters for $Mn_{1.05(1)}Rh_{0.02(1)}Bi$.

| | |
|---|---|
| Formula | $Mn_{1.05(1)}Rh_{0.02(1)}Bi$ |
| Formula weight | 268.18 |
| Crystal system | Orthorhombic |
| Space group; $Z$ | $Fdd2$; 128 |
| Unit cell dimensions (Å) | $a = 8.683(3)$<br>$b = 47.704(8)$<br>$c = 15.021(3)$ |
| Volume (Å$^3$) | 6222(3) |
| Density (calculated) g/cm$^3$ | 9.161 |
| Absorption coefficient (mm$^{-1}$) | 96.742 |
| Crystal size (mm$^3$) | $0.08 \times 0.05 \times 0.03$ |
| $\theta$ range (°) | 2.743–28.981 |
| Index ranges | $-11 \le h \le 11$,<br>$-63 \le k \le 62$,<br>$-20 \le l \le 20$ |
| Reflections collected | 17387 |
| Independent reflections | 3880 [R(int) = 0.0616] |
| Completeness to $\theta = 25.242°$ | 99.9 % |
| Absorption correction | multiscan |
| Refinement method | Full-matrix least-squares on $|F|^2$ |
| Data/restraints/parameters | 3880/1/154 |
| Goodness of fit on $|F|^2$ | 1.089 |
| Final R indices [$I > 2\sigma(I)$] | $R1 = 0.0415$, $wR2 = 0.1096$ |
| R indices (all data) | $R1 = 0.0550$, $wR2 = 0.1184$ |
| Absolute structure parameter | −0.019(14) |
| Extinction coefficient | 0.0000152(7) |
| Largest diff. peak and hole | 5.323 and −3.600 e.Å$^{-3}$ |

TABLE II. Atomic coordinates and equivalent isotropic displacement parameters (Å$^2$) for $Mn_{1.05(1)}Rh_{0.02(1)}Bi$. $U_{eq}$ is defined as one-third of the trace of the orthogonalized $U_{ij}$ tensor.

| Atom | Wyck. | Occ. | $x$ | $y$ | $z$ | $U_{eq}$ | |
|---|---|---|---|---|---|---|---|
| Bi1 | 8$a$ | 1 | 0 | 0 | 0 | 0.013(1) | Layer A |
| Bi2 | 8$a$ | 1 | 0 | 0 | 0.5197(1) | 0.009(1) | |
| Bi3 | 16$b$ | 1 | 0.0292(1) | 0.2501(1) | 0.0291(1) | 0.013(1) | |
| Bi4 | 16$b$ | 1 | 0.2490(1) | 0.1883(1) | 0.4359(1) | 0.010(1) | Layer B |
| Bi5 | 16$b$ | 1 | 0.5005(1) | 0.0586(1) | 0.2016(1) | 0.015(1) | |
| Bi6 | 16$b$ | 1 | 0.0230(1) | 0.3087(1) | 0.1780(1) | 0.015(1) | |
| Bi7 | 16$b$ | 1 | 0.0239(1) | 0.1917(1) | 0.1782(1) | 0.014(1) | |
| Bi8 | 16$b$ | 1 | 0.5009(1) | 0.1247(1) | 0.0188(1) | 0.011(1) | Layer A' |
| Bi9 | 16$b$ | 1 | 0.2489(1) | 0.1250(1) | 0.2696(1) | 0.012(1) | |
| Mn1 | 8$a$ | 0.73(1) | 0 | 0 | 0.1869(1) | 0.010(1) | tbp |
| Rh1 | 8$a$ | 0.27(1) | 0 | 0 | 0.1869(1) | 0.010(1) | |
| Mn2 | 16$b$ | 1 | 0.5004(3) | 0.0630(1) | 0.0192(1) | 0.015(1) | |
| Mn3 | 16$b$ | 1 | 0.2480(3) | 0.0306(1) | 0.1032(1) | 0.013(1) | Layer c |
| Mn4 | 16$b$ | 1 | 0.0021(3) | 0.2194(1) | 0.3537(1) | 0.015(1) | |
| Mn5 | 16$b$ | 1 | 0.2498(2) | 0.2193(1) | 0.6014(1) | 0.012(1) | |
| Mn6 | 16$b$ | 1 | 0.2521(3) | 0.0924(1) | 0.1015(1) | 0.015(1) | Layer c' |
| Mn7 | 16$b$ | 1 | 0.2478(3) | 0.5928(1) | 0.1006(1) | 0.014(1) | |
| Mn8 | 16$b$ | 1 | 0.0019(3) | 0.0927(1) | 0.3543(1) | 0.014(1) | |
| Mn9 | 16$b$ | 1 | 0.2483(3) | 0.1545(1) | 0.1026(1) | 0.015(1) | |

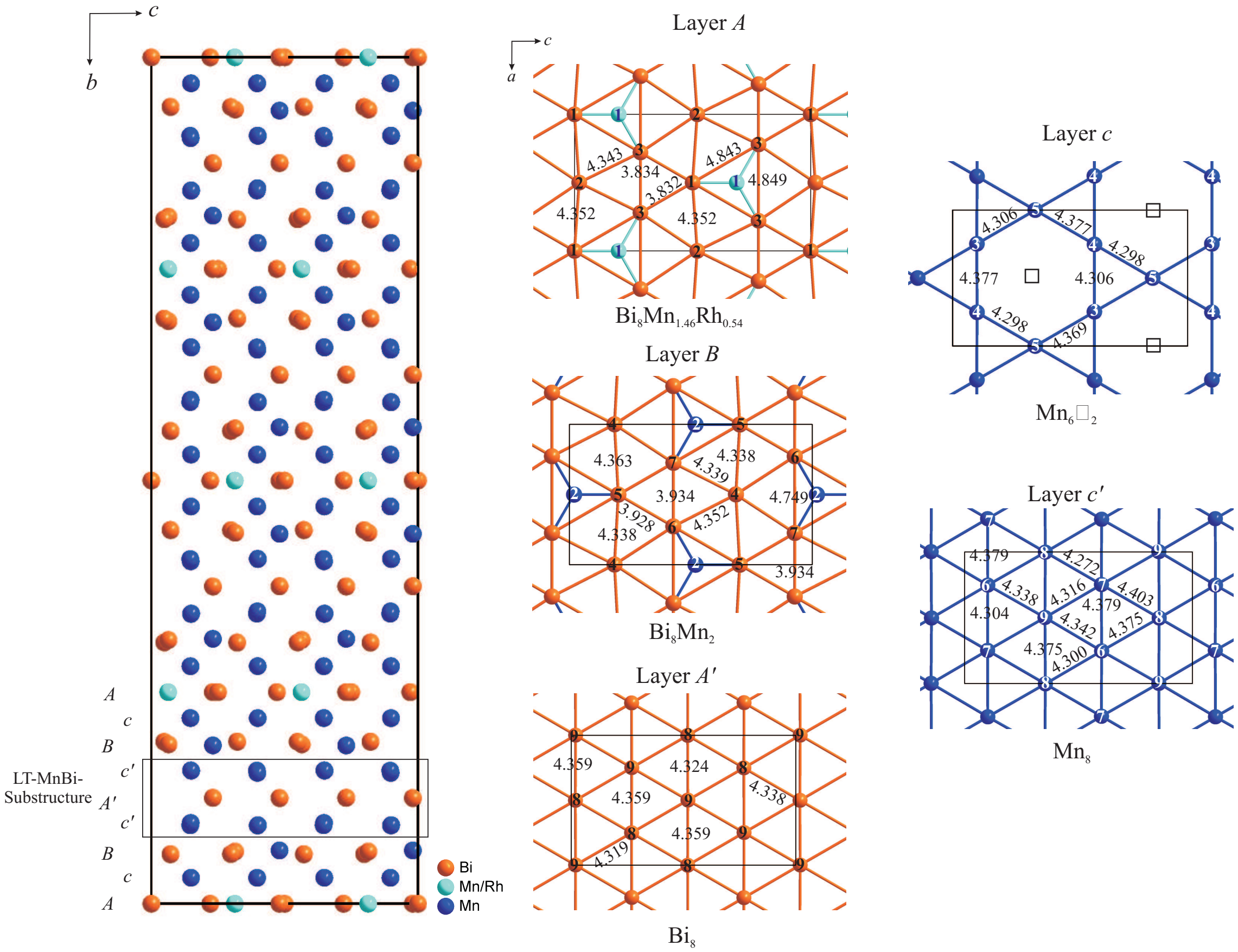

FIG. 10. (Color online)Crystal structure of orthorhombic $Mn_{1.05(1)}Rh_{0.02(1)}Bi$. Unit cell viewed along the $a$ axis (top left). Five individual 2D atomic net along the $b$ axis, show different atomic arrangement and compositions. In layer $c$, Mn vacancies are represented by open squares.